\documentclass[11pt]{article}
\usepackage{epsfig}
\usepackage{dcolumn}
\usepackage{bm}
\usepackage{graphicx}
\usepackage{amssymb,amsmath}
\usepackage{multirow}
\usepackage{cite,color,url}
\usepackage{tabu}
\usepackage{array}
\usepackage[colorlinks=true,urlcolor=blue,anchorcolor=blue
,citecolor=blue,filecolor=blue,linkcolor=blue,menucolor=blue
,linktocpage=true,pdfproducer=medialab,pdfa=true]{hyperref}

\usepackage{colordvi}
\usepackage{epsfig,psfrag,rotating,soul}
\def\beqa{\begin{eqnarray}}
\def\eeqa{\end{eqnarray}}

\newcommand{\ra}{\rightarrow}
\newcommand{\hp}{H^\pm}

\evensidemargin \oddsidemargin
\allowdisplaybreaks

\psfrag{lk}{$l_k$}
\psfrag{lm}{$l_m$}
\psfrag{blk}{$\bar {l}_k$}
\psfrag{blm}{$\bar {l}_m$}
\def\thefootnote{\fnsymbol{footnote}}

\newcommand{\lsim}{\stackrel{\scriptstyle <}{{ }_{\sim}}}
\let\OLDthebibliography\thebibliography
\renewcommand\thebibliography[1]{
\OLDthebibliography{#1}
\setlength{\parskip}{0pt}
\setlength{\itemsep}{0pt plus 0.3ex}}
\usepackage{makecell}

\begin{document}

\thispagestyle{empty}
\begin{center}
\begin{Large}
\textbf{\textsc{New Discovery Modes \\[0.15cm] 
for a Light Charged Higgs Boson at the LHC}}
\end{Large}

\vspace{1cm}
{\sc
A. Arhrib$^{1}$%
\footnote{\tt \href{mailto:aarhrib@gmail.com}{aarhrib@gmail.com}}%
, 
R. Benbrik$^{2}$%
\footnote{\tt
\href{mailto:r.benbrik@uca.ma}{r.benbrik@uca.ma}}
,
M. Krab$^{3}$%
\footnote{\tt
\href{mailto:mohamed.krab@usms.ac.ma}{mohamed.krab@usms.ac.ma}}
,                            
B. Manaut$^{3}$%
\footnote{\tt
\href{mailto:b.manaut@usms.ma}{b.manaut@usms.ma}}
,
S. Moretti$^{4}$%
\footnote{\tt
\href{mailto:s.moretti@soton.ac.uk}{s.moretti@soton.ac.uk}}
,
Yan  Wang$^{5}$%
\footnote{\tt
\href{mailto:wangyan@imnu.edu.cn}{wangyan@imnu.edu.cn}}
,
Qi-Shu Yan$^{6,7}$%
\footnote{\tt
\href{mailto:yanqishu@ucas.ac.cn}{yanqishu@ucas.ac.cn}}
}

\vspace*{.7cm}

{\sl
$^1$ Abdelmalek Essaadi University, Faculty of Sciences and techniques,
Tanger, Morocco.}\\
{\sl
$^2$ Laboratoire  de  Physique  Fondamentale  et  Appliquée  Safi, Facult\'e Polydisciplinaire de Safi,
Sidi Bouzid, B.P. 4162,  Safi, Morocco.}

{\sl
$^3$ Sultan Moulay Slimane University, Polydisciplinary Faculty,
Research Team in Theoretical Physics and Materials (RTTPM), Beni Mellal, 23000, Morocco.}

{\sl
$^4$School of Physics and Astronomy, University of Southampton, Southampton, SO17 1BJ, United Kingdom.}

{\sl
$^5$ College of Physics and Electronic Information, Inner Mongolia Normal University, Hohhot 010022, PR China.}

{\sl
$^6$ Center for Future High Energy Physics, Chinese Academy of Sciences, Beijing 100049, China.}

{\sl $^7$ School  of  Physics  Sciences,  University  of  Chinese  Academy  of  Sciences,  Beijing  100039,  PR  China.} 

\end{center}

\vspace*{0.1cm}

\begin{abstract}
\noindent
		 At the Large Hadron Collider (LHC), both the ATLAS and CMS Collaborations have been searching for light charged Higgs bosons via top (anti)quark production and decays channels, like $pp\to t \bar{t}$ with one top (anti)quark decaying into a charged Higgs boson and a $b$ (anti)quark,  when the decay  is kinematically open (i.e., when $m_{H^\pm}\lsim m_t$). In this paper, we propose new searches at the LHC involving  light charged Higgs bosons via their pair  production channels like $pp\to H^\pm h/A$ and $pp\to H^+ H^-$ 
		in the 2-Higgs Doublet Model (2HDM) Type-I and -X scenarios. By focusing on the case where the heavy $H$ state plays the role of the Standard Model (SM)-like Higgs boson with a mass near 125 GeV, we study the aforementioned Higgs boson pair production channels and investigate their bosonic decays, such as $H^\pm \to W^{\pm  } h$ and/or $H^\pm \to W^{\pm  } A$. We demonstrate that for a light charged Higgs boson state, with $m_{H^\pm}\lsim m_t$,  at the LHC, such di-Higgs production and decay channels can give rise to signatures with event rates much larger than those emerging from $pp\to t\bar
		t\to t\bar b H^-$ + c.c.. We specifically study $h/A\to b\bar b$ and $\tau^+\tau^-$ decays. We, therefore, claim that the discussed combination of new  production and decay modes can result in an alternative discovery channel for charged Higgs bosons lighter than the top (anti)quark at the LHC within the above two 2HDM Types. Finally, 
		in order to motivate experimentalists in ATLAS and CMS to search for such signatures, we propose 16 Benchmark Points (BPs) which are compatible with both theoretical and experimental constraints.
\end{abstract}
 
\def\thefootnote{\arabic{footnote}}
\setcounter{page}{0}
\setcounter{footnote}{0}
\newpage
\section{Introduction}
\label{intro}
Following the discovery of a 125 GeV Higgs boson in the first 
run of the LHC \cite{Aad:2012tfa,Chatrchyan:2012ufa}, 
several studies of its properties were undertaken. The 
current situation is that the measured Higgs signal rates in all production and decay 
channels agree  with the SM predictions at the
$\sim2\sigma$ level~\cite{Aad:2019mbh, Sirunyan:2018koj}. 
However, although the current LHC Higgs data are consistent with the SM, 
there is still the possibility that the
observed Higgs state could be part of a model with an extended 
Higgs sector including, e.g., an  extra doublet, singlet and/or triplet. 
Such a possibility is understandable since most of these extended Higgs sector scenarios  possess a decoupling limit and give back the SM after integrating out the heavier states.
%
Typically, most of the higher Higgs representations with an extra doublet
predict one or more charged Higgs bosons. Therefore, 
it will be a smoking gun for  new physics if a  charged Higgs bosons were to be found. Indeed, one of 
the main goals of future LHC runs, in addition to raising the precision of the existing 
measurements of the discovered Higgs boson properties, is to pursue direct searches for new 
Higgs states in the quest to discover the evidence of new physics.

The 2-Higgs Doublet Model (2HDM) is one of the simplest extensions of the SM. It contains
 two Higgs doublets, labelled as $\Phi_1$ and $\Phi_2$, which can generate masses to all fermions and gauge bosons. The particle spectrum of the 
2HDM  includes two CP even Higgs bosons ($h$ and $H$, with $m_h<m_H$), 
one CP odd Higgs boson ($A$)  and a pair of charged   Higgs
bosons ($H^\pm$)\footnote{Herein, we assume that the $H$ state of the 2HDM is the discovered SM-like Higgs boson state.}.  Within such a framework,
at hadron colliders, a charged Higgs boson can be produced through several  channels. In particular, a  light charged Higgs boson (with $m_{H\pm} \leq m_t-m_b$)  can be copiously produced from the $pp\to t\bar{t}$ process via the top decay $t\to bH^+$ (or the equivalent antitop mode).
When kinematically allowed, 
$pp\to t\bar{t}\to b\bar{b}H^- W^+ $  + c.c. provides the largest production rate for a light charged Higgs bosons. 
However, other production mechanisms have been studied in the literature,  listed below.
\begin{itemize}
\item Single production: $gb \to tH^-$ + c.c. and $gg\to t\bar{b}H^-$, the latter containing the $t\bar t$ production and decay mode, appropriately combined \cite{Barger:1993th,Gunion:1986pe,Barnett:1987jw,DiazCruz:1992gg,Moretti:1999bw,Miller:1999bm,Guchait:2001pi,Alwall:2003tc,Alwall:2004xw}.
\item Associated production with a $W^\pm$ gauge boson: $gg \to W^\pm H^\mp $ and
$b\bar{b} \to  W^\pm H^\mp $ \cite{BarrientosBendezu:1998gd,Moretti:1998xq,BarrientosBendezu:1999vd}.

 \item Production through $W^\pm b$  scattering: $qb\to q' H^+ b$  \cite{Moretti:1996ra,Arhrib:2015gra}.

\item Resonant production via $c \bar s, c\bar b\to H^+$  \cite{HernandezSanchez:2012eg,Hernandez-Sanchez:2020vax,Dittmaier:2007uw}.

\item Associate production with a neutral Higgs: $q\bar q' \to  H^\pm\phi$ where 
$\phi$ denotes one of the  three neutral 
 Higgs bosons, $\phi=h,H$ or $A$ \cite{cpyuan}  (see also Refs.~\cite{Enberg:2014pua,Enberg:2015qsa,Enberg:2017gyo,Enberg:2018nfv,Enberg:2018pye}).
 
\item Pair production: $gg\to H^+ H^-$ and $q\bar q\to H^+ H^-$ 
 \cite{BarrientosBendezu:1999gp,Brein:1999sy,
Moretti:2001pp,Moretti:2003px}.

\end{itemize}
(See also Refs.~\cite{Aoki:2011wd,Akeroyd:2016ymd,Arhrib:2018ewj,Coleppa:2019cul} for a review of all available $H^\pm$ hadro-production modes in 2HDMs.)

Before the opening of the $t\bar{b}$ channel, a charged Higgs boson decays predominantly in 
$\tau\nu$ followed by $c\bar{s}$, with $c\bar{b}$ being Cabibbo-Kobayashi-Maskawa (CKM) suppressed.  However, it has been shown that, 
in the presence of a  light CP-even, $h$, or CP-odd, $A$, Higgs boson, the bosonic decays of the charged Higgs bosons $H^\pm\to W^\pm h/A$ would compete with the  $\tau\nu$ and $c\bar{s}$ modes and could even be dominant in some cases \cite{Akeroyd:1998dt,Arhrib:2016wpw}.  Once the $t\bar{b}$ channel is open, $\tau\nu$ and  $c\bar{s}$ 
would be suppressed, so that $t\bar{b}$ would compete with the bosonic decay channels 
such as $H^\pm\to W^\pm h/A$.

A charged Higgs boson has been searched for at the Tevatron through top decays,  $pp\to t\bar{t}\to b \bar{b}H^- W^+$,  
 followed by  either  $\tau\nu$ or $c\bar{s}$ decays. Negative searches have been used to set a limit on BR$(t\to bH^+)\times {\rm BR}(H^+\to \tau \nu)$ \cite{Abazov:2009ae,Abulencia:2005jd}. Alongside Tevatron searches, charged Higgs bosons have been 
searched for also at LEP  using $e^+e^-\to \gamma^*, Z^*\to H^+ H^-$  followed by either $H^\pm \to \tau \nu$, 
$H^\pm \to c{s}$ or $H^\pm \to W^\pm A$ \cite{Abbiendi:2013hk}. If the charged Higgs boson decays  dominantly to $\tau\nu$ or $c\bar{s}$,  LEP2 sets a 
lower bound on the mass of the order of 80 GeV while in the case where charged Higgs  decays are dominated by $W^{\pm}A$, via a  light CP-odd Higgs state ($m_{A}\approx 12$ GeV), the lower bound on the charged Higgs  mass is about 72 GeV \cite{Abbiendi:2013hk}. 
At the LHC, both ATLAS and CMS have been searching for charged Higgs bosons  
either from top decay, in case $t\to b H^+ $ is open, or from the first production mechanisms listed above, $pp\to tH^-$ or $pp \to t\bar{b}H^-$, otherwise.
For a light charged Higgs,  ATLAS and CMS experiments have already set  an exclusion on the product of Branching Ratios (BRs), 
BR$(t\to bH^+) \times {\rm BR}(H^+ \to \tau^+ \nu)$ 
 \cite{Aad:2014kga,Khachatryan:2015qxa,CMS:2016szv,Aaboud:2018gjj}, of order percent in the mass range of 80 to 160 GeV.
 Other channels, such as  
$H^+ \to c\bar{s}$, have also been searched for by ATLAS and CMS.
Assuming that BR$(H^+ \to c\bar{s})=100\%$, one can set a limit on BR$(t\to bH^+)$  in the range $1.68\%$ to $0.25\%$ for a charged Higgs boson mass between 90 and 160 GeV  \cite{ATLAS:2011yia,Sirunyan:2020aln,Tanabashi}.
In the case of a heavy charged Higgs boson, ATLAS and CMS provide a limit 
on cross section times BR$(H^\pm\to \tau \nu_\tau)$ or   
BR$(H^+\to tb)$ \cite{Sirunyan:2019hkq,Aad:2014kga,Khachatryan:2015qxa} (but these are not relevant here).  

The aim of this letter is to revisit the last two $H^\pm$ production modes listed above in the context of the upcoming LHC Run 3, to assess to what extent 
they could supplement $H^\pm$ searches via $pp\to t\bar t\to b\bar b W^+H^-$ + c.c.. We will 
show that the production rates of such alternative di-Higgs production channels,  via $pp\to H^\pm h/A$ and $pp\to H^+ H^-$,
 can be overwhelmingly larger than those of the top quark associated production channels. Specifically, we will show that the bosonic decays of a {light} 
charged Higgs boson, via $H^\pm \to W^{\pm }h/A$, could be dominant and lead to accessible signatures alternative   to those emerging from the top-antitop production and decay.  Hence, such modes may serve as new discovery channels for light $H^\pm$ states at the LHC. We further stress that, just like $pp\to t\bar t$, also $pp\to H^\pm h/A$ and $pp\to H^+ H^-$ are model independent as they are predominantly mediated by $s$-shannel gauge boson exchange with no 2HDM parameters involved\footnote{In fact, even the $pp\to H^\pm h$ channel is effectively model independent because of the properties of the SM-like Higgs boson discovered at the LHC, as we shall discuss later.}.

The paper is organised as follows. In Section 2 we give a brief review of the 2HDM 
 and the Yukawa couplings used. In Section 3 we list the theoretical and experimental 
 constraints that will be used during our study. Our numerical results and a set of Benchmark Points (BPs) are given in Sections 4 and 5, respectively. Our conclusion 
is given in Section 6.
 
\section{A Review of the 2HDM}

The general 2HDM is obtained by extending the SM Higgs sector, which has  one doublet scalar field $\Phi_1$ with $Y= +1$, 
with an additional doublet scalar field $\Phi_2$ with $Y = +1$. The most
general renormalisable potential which is invariant under 
$\rm{SU(2)_L \times U(1)_Y}$ is given by \cite{Branco:2011iw}:
\begin{eqnarray}
 V_{\rm{Higgs}}(\Phi_1,\Phi_2) &=& \lambda_1(\Phi_1^\dagger\Phi_1)^2 +
\lambda_2(\Phi_2^\dagger\Phi_2)^2 +
\lambda_3(\Phi_1^\dagger\Phi_1)(\Phi_2^\dagger\Phi_2) +
\lambda_4(\Phi_1^\dagger\Phi_2)(\Phi_2^\dagger\Phi_1) \nonumber  +~\nonumber\\ && +
\frac12\left[\lambda_5(\Phi_1^\dagger\Phi_2)^2 +\rm{h.c.}\right]
+~\left\{\left[\lambda_6(\Phi_1^\dagger\Phi_1)+\lambda_7(\Phi_2^\dagger\Phi_)\right]
(\Phi_1^\dagger\Phi_2)+\rm{h.c.}\right\} \nonumber \\ && 
-~\left\{m_{11}^2 \Phi_1^\dagger \Phi_1+ m_{22}^2\Phi_2^\dagger
\Phi_2 + \left[m_{12}^2
\Phi_1^\dagger \Phi_2 + \rm{h.c.}\right] \right\}.\label{CTHDMpot}
\end{eqnarray}
By the hermiticity of Eq. (\ref{CTHDMpot}),
$\lambda_{1,2,3,4}$ as well as $m_{11}^2$ and $m_{22}^2$ are real-valued 
while the parameters $\lambda_{5}$, $\lambda_6$, $\lambda_7$ and $m_{12}^2$ are in general complex and can generate CP violation in the Higgs sector.
After Electro-Weak Symmetry Breaking (EWSB) takes place, of the 8 degrees of freedom initially present in $\Phi_1$ and $\Phi_2$, 3 correspond to the longitudinal components of $W^{\pm}$ and $Z^0$ while  
the remaining 5 appear as the physical Higgs fields mentioned in the previous section: 
$h$, $H$,  $A$ and $H^{\pm}$. 

We finally end up with 7 free parameters in the 2HDM, which here we choose to be 
\begin{eqnarray}
m_{h}, \,m_{H}, \,m_{A}, \,m_{H^{\pm}}, \,\alpha, \,\beta\ \,{\rm and} \ \,m_{12}^2.
\label{eq:param} 
\end{eqnarray}
One of the CP-even Higgs bosons must be the one discovered at the LHC in 2012 (as intimated, we assume this to be the $H$ state). Furthermore, in the list above, $\alpha$ is the mixing angle between the two CP-even Higgs bosons while $\tan\beta$ is the ratio of the Vacuum Expectation Values (VEVs) of the two Higgs doublet fields.

It is well known that the presence of a single doublet in the SM plays a dual role, of generating the mass of both the gauge bosons and fermions (the latter via the introduction of Yukawa couplings). This dynamic produces two strong constraints. The first constraint is 
linked to the bosonic sector: $\rho= m_{W^\pm}^2/(m_Z^2\cos\theta_W^2) \approx 1 $  which is in good agreement with
experiments. The second constraint is linked to the fermionic sector and 
guarantees the absence of Flavour Changing Neutral Currents (FCNCs). 
So, any extension of the SM must be done by preserving these two constraints. 
In the framework of the 2HDM, the first constraint is satisfied because of the 
doublet representation of the Higgs fields, so long certain Higgs mass relations are respected. In the Yukawa sector, though, with the presence of two Higgs doublets, though, if we allow for EWSB like in the SM, 
we end up with FCNCs in the Yukawa sector at the tree level already.
To avoid the latter, the most elegant solution is dictated by the Glashow-Weinberg theorem \cite{Glashow:1976nt}: FCNCs can be avoided at tree level
 if we impose a discrete $Z_2$ symmetry such that  each fermion type couples only  to one of the Higgs doublets. Thus, there are four Types of 2HDM
\cite{Branco:2011iw}. In this study, though, we are interested only in two of these, Type-I and -X.
 In the 2HDM Type-I, only the doublet $\Phi_2$ couples to all the fermions exactly like 
  in the SM while in the 2HDM Type-X (or lepton-specific) all the quarks couple 
  to $\Phi_2$ and  the charged leptons couple to $\Phi_1$. If we extend such a $Z_2$ symmetry to the scalar potential, wherein $\Phi_1\to + \Phi_1$ and
$\Phi_2\to - \Phi_2$, this requires 
  $\lambda_6=\lambda_7=0$. Note that, however, the $m_{12}^2$ term therein, being of dimension-2,  can remain, as it would only break the $Z_2$ symmetry softly, thereby allowing for FCNCs compliant with experimental constraints. 

The neutral and charged Higgs couplings to fermions can be obtained from the Yukawa Lagrangian and are given by \cite{Branco:2011iw}
\begin{eqnarray}
 - {\mathcal{L}}_{\rm Yukawa} = \sum_{f=u,d,l} \left(\frac{m_f}{v} \kappa_f^h \bar{f} f h + 
 \frac{m_f}{v}\kappa_f^H \bar{f} f H 
 - i \frac{m_f}{v} \kappa_f^A \bar{f} \gamma_5 f A \right) + \nonumber \\
 \left(\frac{V_{ud}}{\sqrt{2} v} \bar{u} (m_u \kappa_u^A P_L +
 m_d \kappa_d^A P_R) d H^+ + \frac{ m_l \kappa_l^A}{\sqrt{2} v} \bar{\nu}_L l_R H^+ + H.c. \right),
 \label{Yukawa-1}
\end{eqnarray}
where $\kappa_f^S$ are the Yukawa couplings in the 2HDM, which are listed 
in Table \ref{Yukawa-2} for the two Types of interest here. Here, $V_{ud}$ represents a CKM matrix element. 

\begin{table}
 \begin{center}
  \begin{tabular}{||l|l|l|l|l|l|l|l|l|l||}
   \hline
    & $\kappa_u^h$ & $\kappa_d^h$ & $\kappa_l^h$ & $\kappa_u^H$ & $\kappa_d^H$ & $\kappa_l^H$ & $\kappa_u^A$ & $\kappa_d^A$ & $\kappa_l^A$ \\ \hline
    Type-I & $c_\alpha/s_\beta$ & $c_\alpha/s_\beta$& $c_\alpha/s_\beta$ & $s_\alpha/s_\beta$ & $s_\alpha/s_\beta$ & $s_\alpha/s_\beta$ & $c_\beta/s_\beta$ & 
    $-c_\beta/s_\beta$ & $-c_\beta/s_\beta$ \\ \hline
    Type-X & $c_\alpha/s_\beta$ & $c_\alpha/s_\beta$& $-s_\alpha/c_\beta$ & $s_\alpha/s_\beta$ & $s_\alpha/s_\beta$ & $c_\alpha/c_\beta$ & $c_\beta/s_\beta$ & 
    $-c_\beta/s_\beta$ & $s_\beta/c_\beta$ \\ \hline
    \end{tabular}
 \end{center}
 \caption{Higgs couplings to fermions in the 2HDM Type-I and -X.}
\label{Yukawa-2}
\end{table}

As one can read from such a  table, the Higgs couplings to quarks are identical in 2HDM Type-I and -X, the only difference being the coupling to leptons.
From the phenomenological point of view, the most relevant features are that, in the 2HDM Type-X, the Higgs couplings to leptons enjoy an enhancement for large $\tan\beta$ while, in the 2HDM Type-I, the couplings to all fermions are suppressed if $\tan\beta \gg 1$. 

\section{Theoretical and Experimental Constraints}
The 2HDM parameter space is limited by both theoretical and experimental constraints.  The theoretical constraints that have been imposed on the scalar potential are as follows. 
\begin{itemize}
\item Vacuum stability conditions, which require the Higgs potential to be bounded from below \cite{sta}, are given by the following inequalities:
	\begin{equation}
	\lambda_1 > 0,\quad \lambda_2 > 0,\quad \lambda_3 > - \sqrt{\lambda_1 \lambda_2},\quad \lambda_3 + \lambda_4 - \mid\lambda_5\mid > -\sqrt{\lambda_1 \lambda_2}.
	\end{equation}

\item Perturbativity constraints imply that all quartic coefficients of the scalar potential satisfy the condition $\mid\lambda_i\mid \leq 8\pi$ ($i=1,... 5$).

\item Perturbative unitarity constraints \cite{uni1,uni2} require that scattering processes involving gauge bosons and Higgs states remain unitary at high energy.    
\end{itemize}
We also take into account experimental constraints from direct Higgs boson searches at  
 LEP, Tevatron, LHC as well as Electroweak Precision Observables 
 (EWPOs) plus flavour physics data are also used. 
 
\begin{itemize}
\item Measurements of  the oblique parameters $S$, $T$ and $U$ \cite{peskin} 
	can constrain several new physics models and, in the case of the 2HDM, they can be used to set a limit on the mass splitting between the physical Higgs bosons (specifically, a charged and a neutral Higgs state) and on the mixing angles as well. The 2HDM contributions to $S$, $T$ and $U$ parameters are taken from Refs. \cite{stu-2HDM1,stu-2HDM2,stu-2HDM3}.
	
\item Exclusion limits at $95\%$ Confidence Level (CL) from Higgs searches at colliders (LEP, Tevatron and LHC) via \texttt{HiggsBounds-5.9.0} \cite{HB} and 
compliance with  SM-like  Higgs  boson measurements via \texttt{HiggsSignals-2.6.0} \cite{HS} are enforced.
	
\item B-physics observables are calculated using the public code \texttt{SuperIso v4.1} \cite{superIso}. Specifically, we have used the following observables:
\begin{enumerate}
	\item BR$(B \to X_s \gamma) = (3.32 \pm 0.15) \times 10^{-4}$ \cite{Bphys1},
	\item BR$(B_s \to \mu^+ \mu^-) = (3.0 \pm 0.6 \pm 0.25) \times 10^{-9}$ \cite{Bphys2},
	\item BR$(B_u \to \tau \nu) = (1.06 \pm 0.19) \times 10^{-4}$ \cite{Bphys1}. 
\end{enumerate}

\end{itemize}
Note that we have used the public code \texttt{2HDMC} \cite{2HDMC} to check the theoretical constraints such as perturbative unitarity,  perturbativity of the $\lambda_i$, 
vacuum stability of the scalar potential as well as the oblique parameters 
$S$, $T$ and $U$.  

\section{Numerical Results}
As previously mentioned, in what follows, we assume that the SM-like Higgs particle discovered at CERN in 2012 is the $H$ state, for which $m_H=125$ GeV, so that $m_h$ would be smaller in comparison. We then perform a systematic scan over the 2HDM parameter space as indicated in Table \ref{param_scan}, for both Type-I and -X.
\begin{table}[hbtp]
	\centering
	\begin{tabular}{|l|l|} \hline\hline
		Parameters &  \quad2HDM-I, -X  \\ \hline
		~~\quad $m_h$  & ~~\,\quad$[10,\,120]$  \\
		~~\quad $m_H$ &  ~~\,\qquad$125$  \\
		~~\quad $m_A$ &  ~~\,\quad$[10,\,120]$  \\  
		~~\quad $m_{H^\pm}$ &  ~~\,\quad$[80,\,170]$  \\ 
		~~\quad $s_{\beta-\alpha}$ & ~~\,$[-0.3,\,-0.05]$ \\ 
		\,~\quad $\tan\beta$ & ~\qquad$[2,\,60]$  \\
		\,~~\quad $m_{12}^2$ & $[0,\,m^{2}_H \sin\beta \cos\beta]$  \\
		\quad $\lambda_6=\lambda_7$ & ~\quad\qquad$0$  \\ \hline\hline	  
	\end{tabular}	
	\caption{2HDM Type-I and -X parameter scans adopted (all masses are in GeV).} \label{param_scan}
\end{table}

Since we assume that $m_H=125$ GeV, then the $S$, $T$ and $U$ constraints will force the whole Higgs boson  
spectrum of the 2HDM to be rather light. The charged Higgs is taken in the range $80-170$ GeV and the CP-odd is presumed to be in the range $10-120$ GeV.
Having assumed that $H$ is the known Higgs particle, taking into account all LHC data
will force the $H$ couplings to  SM particles to be rather SM-like. 
As a consequence, the coupling of the $H$ state to gauge bosons $W^+W^-$ (and $ZZ$), which is 
proportional to  $\cos(\beta-\alpha)$, would be SM-like if $\cos(\beta-\alpha)\approx 1$. This trigonometric function is the same entering the coupling $W^+H^-h$, which is then essentially independent of the $\alpha$ and $\beta$ parameters of the 2HDM, as already remarked (just like $W^+H^-A$).   

We are interested in light charged Higgs bosons, which could come 
either from (anti)top decay or Higgs pair production: i.e., 
$gg, q\bar{q} \to t\bar{b}H^-$ + c.c., $q\bar{q} \to H^+ H^-$, 
and $q\bar{q}' \to H^+ h/A$ + c.c.\footnote{Therefore, in our approach, we privilege the use of the $2\to3$ description of the first channel, i.e., in the 4-Flavour Scheme (4FS), as it is the most suitable to model the so-called `transition region' $m_{H^\pm}\approx m_t$ \cite{Guchait:2001pi,Assamagan:2004gv}. However,
as it is customary in many experimental analyses, we will also present results for the process $pp\to t\bar t$ followed by $t\to bH^+$ (and c.c.), wherein the (anti)top quark is treated in Narrow Width Approximation (NWA), again, in LO approximation.}.
We compute cross sections at Leading Order (LO) only (i.e., at tree level), though Next-to-LO (NLO) corrections from Quantum Chromo-Dynamics (QCD) exist for all processes: see Refs. \cite{Zhu:2001nt,Plehn:2002vy,Berger:2003sm} for the top (anti)quark process while those for Higgs pair production can be accounted for through to the use of
NLO Parton Distribution Functions (PDFs). The former (30\%) is somewhat larger than the latter (20\%) for a light $H^\pm$ state, so such different relative effects from QCD should be borne in mind while comparing the two types of processes, but their inclusion will not change our conclusions. 
 
\begin{figure}[hbtp]
\centering
\includegraphics[scale=0.65]{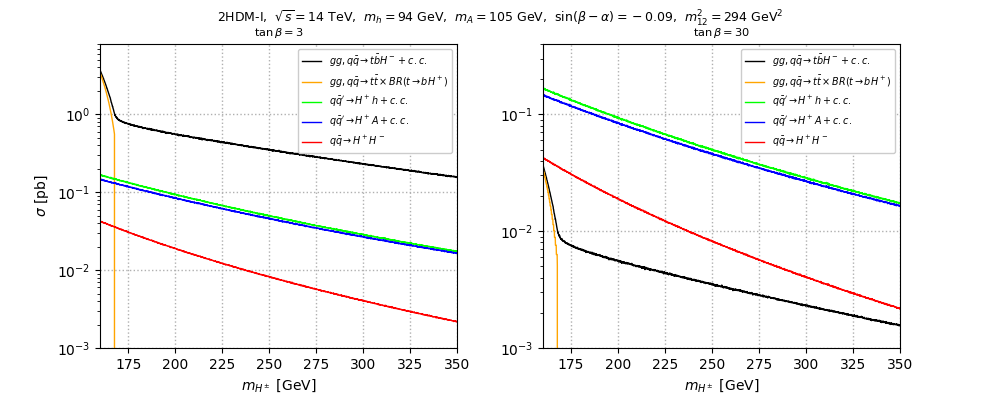}
\caption{Cross section for $gg, q\bar{q} \to t\bar{b}H^-$ + c.c. (black), $gg, q\bar{q} \to t\bar{t} \times {\rm BR}(t \to bH^+)$ + c.c. (orange), $q\bar{q} \to H^+ H^-$ (red) and $q\bar{q}' \to H^\pm\, h/A$ (green/blue) at $\sqrt{s}=14$ TeV, as a function of $m_{H^\pm}$ for $\tan\beta=3$ (left) and $30$ (right). All cross sections are calculated using the code \texttt{MadGraph} \cite{Stelzer:1994ta} at LO. Note that we have used the \texttt{MSTW2008} PDF set from \cite{PlothowBesch:1992qj} with factorisation/renormalisation scale set to $\sqrt{\hat{s}}$, i.e., the centre-of-mass energy at parton level.} \label{fig1}
\end{figure}

We now examine the production rates of light charged Higgs boson at the LHC with $\sqrt s=14$ TeV  from the above  processes. This is done for the 2HDM Type-I  in Figure~\ref{fig1}, over the 
range 160 GeV $\lsim m_{H^\pm}\lsim $ 350 GeV\footnote{This is done for illustrative purposes, though, as the maximum allowed value for $m_{H^\pm}$ in the 2HDM Type-I and -X setups is actually 197 GeV since, for higher values, one would no longer satisfy $S$, $T$ and $U$ constraints.}, for the following values of the other BSM parameters: $m_h=94$ GeV, $m_A=105$ GeV, $\sin(\beta-\alpha)=-0.09$ and $m_{12}^2=294$ GeV$^2$. Here, the left(right) frame refers to $\tan\beta=3(30)$. 
 Let us start with the rates for $\sigma(q\bar q,gg\to t\bar t)$ $\times$ BR($t\to bH^+$) +  c.c., which are clearly only appearing for $m_{H^\pm}\leq m_t-m_b$  and coincide with those for $\sigma(q\bar q,gg\to t\bar bH^-$ +  c.c.) in the limit $m_{H^\pm}\ll 
m_t$.
 The latter, of course, extends beyond such a transition region, though the absence of any resonant contribution in this (heavy) mass region reflects a significant drop in cross section.  
Clearly, as shown in literature and as the plots in Figure~\ref{fig1} demonstrate, the $pp\to  t\bar bH^-$ + c.c. process is important for charged Higgs boson phenomenology at the LHC. However, it is worth noticing that  the  BR$(t \ra b H^+)$ can only be really significant in a parameter space region with small $\tan \beta$ due to the enhancement of the coupling $H^+ \bar t b$ in this model configuration.  As can be seen from Figure~\ref{fig1} (left panel), 2HDM Type-I predicts a scenario where, for $m_{H^\pm}\ll m_t$,  $\sigma(q\bar q,gg\to t\bar{t})\times {\rm BR}(t\to b H^+)$ + c.c.  (and $\sigma(q\bar q,gg\to t\bar b H^-~+~ {\rm c.c.}))$ could reach  values around  30 pb and, in fact, even for very large charged Higgs masses, the rates for the $pp\to t \bar{b} H^-$ + c.c. channel 
remain larger than those for $pp \to \hp h$,  $pp \to \hp A$ as well as $pp \to H^+ H^-$. 
However, as demonstrated in the right panel of Figure \ref{fig1}, when the parameter $\tan\beta$ is large, the top (anti)quark  channels are suppressed by the Yukawa couplings, so that the Higgs pair production channels can largely dominate the $H^\pm$ phenomenology, no matter the value of its mass.  Quite interestingly, once can then exploit  the decay channels $\hp \ra W^{\pm  }\phi$, with $\phi=h$ or $A$. Finally, being the neutral Higgs bosons emerging therein rather light, it is also noted that they finally dominantly decay into two fermions, either $b$'s or $\tau$'s.

With this in mind, 
in Table \ref{tab:sign}, we tabulate the typical final states (left column) and corresponding production and decay patterns (right column) for top pair production and decay, charged Higgs boson associated production with top plus bottom and di-Higgs processes. Notice that the first two channels always have $2W$ bosons in the final states, similarly to the  charged Higgs boson pair production channel. In contrast, the di-Higgs production channels $pp \to H^\pm A/h$ lead to only a single  $W$ in the final states. Below, we will focus on three types of final states, the $4b$ final states (including $2 W + 4b$ and $W +4b$), the $2b2\tau$ final states (including $2 W + 2b 2\tau$ and $W +2b 2 \tau$) and the $4\tau$ final state (including only $W +4 \tau$). 

\begin{table}[hbtp]
	\centering
	\begin{tabular}{|l|l|} \hline\hline
		& Top pair production and  decay chain  \\ \hline
		$\sigma^{h_i}_{2t}(2 W + 2 b + 2f)$  &  2 $\sigma_{t\bar{t}} \times {\rm BR}(t \to b H^+)\times {\rm BR}(\bar{t} \to \bar{b} W^-)\times {\rm BR}(H^\pm \to W^\pm h_i)\; \times $ \\
		& ${\rm BR}(h_i \to f\bar{f})$ \\ 
		\hline \hline 
		& Associated production with top plus bottom and decay chain \\ \hline
		$\sigma^{h_i}_{t}(2 W + 2b + 2f)$  & $ \sigma(pp \to t\bar{b}H^-)\times {\rm BR}(t \to b W^+)\times {\rm BR}(H^\pm \to W^\pm h_i)\; \times $ \\
		& $ {\rm BR}(h_i \to f\bar{f})$ \\
		\hline \hline
		& Di-Higgs production   and  decay chain \\ \hline
		$\sigma^{h_i}_{h_j}(2W+2f + 2f^\prime)$ & $ \frac{1}{1+\delta_{ff'}}$ $\sigma(H^+ H^-)\times {\rm BR}(H^\pm \to W^\pm h_i)\times {\rm BR}(H^\pm \to W^\pm h_j)\; \times $ \\
		&  (${\rm BR}(h_i \to f\bar{f})\times {\rm BR}(h_j \to f^\prime \bar{f}^\prime)$ + BR$(h_j \to f\bar{f})\times {\rm BR}(h_i \to f^\prime \bar{f}^\prime)$)   \\ \hline
		$\sigma^{h_i}_{h_j}(W+2f + 2 f^\prime)$ &$ \frac{1}{1+\delta_{ff'}}$ $\sigma(H^\pm h_i)\times {\rm BR}(H^\pm \to W^\pm h_j)\; \times $ \\
		&  (${\rm BR}(h_i \to f\bar{f})\times {\rm BR}(h_j \to f^\prime \bar{f}^\prime)$ + BR$(h_j \to f\bar{f})\times {\rm BR}(h_i \to f^\prime \bar{f}^\prime)$)  \\
		\hline\hline 
	\end{tabular}
	\caption{The discussed  production processes of charged Higgs bosons are shown here, where the main decay chains which can lead to the given final states are also provided. Here, we define $i,j =1, 2$ and  have $h_1=h$ and $h_2=A$. The symbol $f$($f^\prime$) denotes fermions, like $b$ or $\tau$.} \label{tab:sign}
\end{table}
\begin{figure}[hbtp]
	\centering
	\includegraphics[scale=0.65]{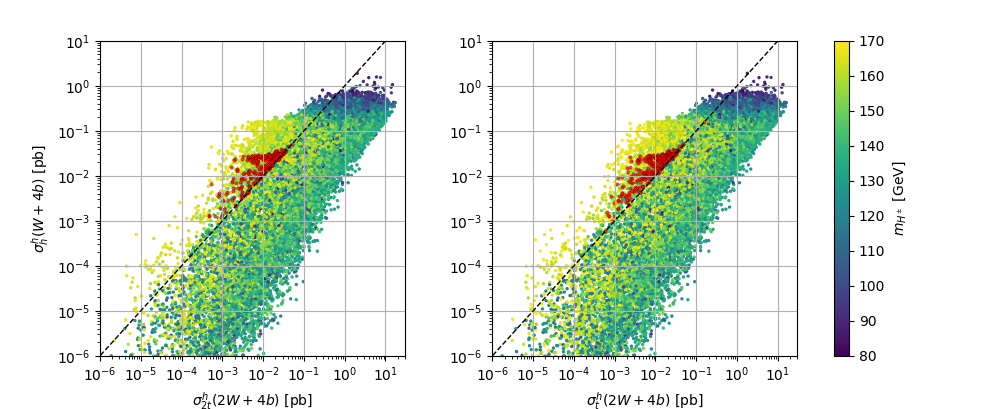}
	\caption{Values of  $\sigma(pp \to H^\pm h)\times {\rm BR}(H^\pm \to W^\pm h)\times {\rm BR}(h \to b\bar{b})^2$ are compared with those of 
		2 $\sigma_{t\bar{t}} \times {\rm BR}(t \to b H^+)\times {\rm BR}(\bar{t} \to \bar{b} W^-)\times {\rm BR}(H^\pm \to W^\pm h)\; \times $ 
		${\rm BR}(h \to b\bar{b})$ 
		(left) and 
		$ \sigma(pp \to t\bar{b}H^-)\times {\rm BR}(t \to b W^+)\times {\rm BR}(H^\pm \to W^\pm h)\; \times $ 
		$ {\rm BR}(h \to b\bar{b})$ 
		(right). The red points identify the values of 
		$\sigma(H^+ H^-)\times {\rm BR}(H^\pm \to W^\pm h)^2 \times {\rm BR}(h \to b\bar b)^2$
		which also exceeds those of the latter two top mediated processes, respectively. 
		The colour bar denotes the mass of the charged Higgs boson. Results are for the 2HDM Type-I.} \label{fig2}
\end{figure}

\begin{figure}[hbtp]
	\centering
	\includegraphics[scale=0.65]{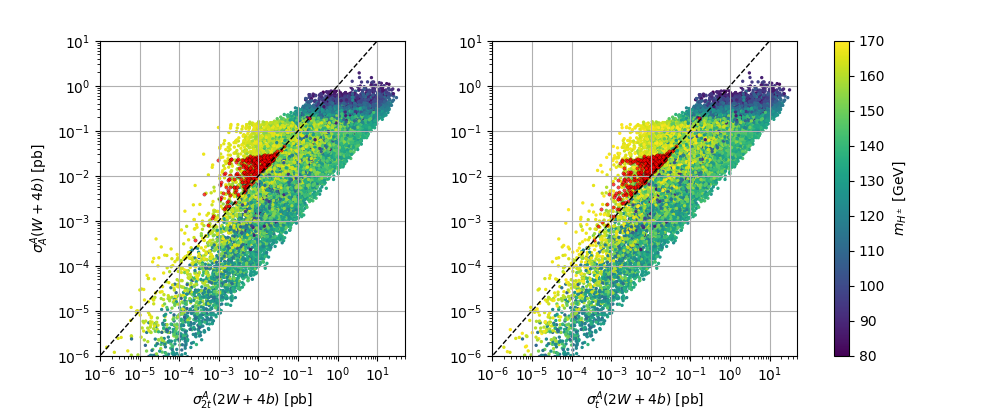}
	\caption{Values of  $\sigma(pp \to H^\pm A)\times {\rm BR}(H^\pm \to W^\pm A)\times {\rm BR}(A \to b\bar{b})^2$ are compared with those of 
		2 $\sigma_{t\bar{t}} \times {\rm BR}(t \to b H^+)\times {\rm BR}(\bar{t} \to \bar{b} W^-)\times {\rm BR}(H^\pm \to W^\pm A)\; \times $ 
		${\rm BR}(A \to b\bar{b})$ 
		(left) and 
		$ \sigma(pp \to t\bar{b}H^-)\times {\rm BR}(t \to b W^+)\times {\rm BR}(H^\pm \to W^\pm A)\; \times $ 
		$ {\rm BR}(A \to b\bar{b})$ 
		(right). The red points identify the values of 
		$\sigma(H^+ H^-)\times {\rm BR}(H^\pm \to W^\pm A)^2 \times {\rm BR}(A \to b\bar b)^2$
		which also exceeds those of the latter two top mediated processes, respectively. 
		The colour bar denotes the mass of the charged Higgs boson. Results are for the 2HDM Type-I.} \label{fig3}
\end{figure}

\begin{figure}[hbtp]
	\centering
	\includegraphics[scale=0.65]{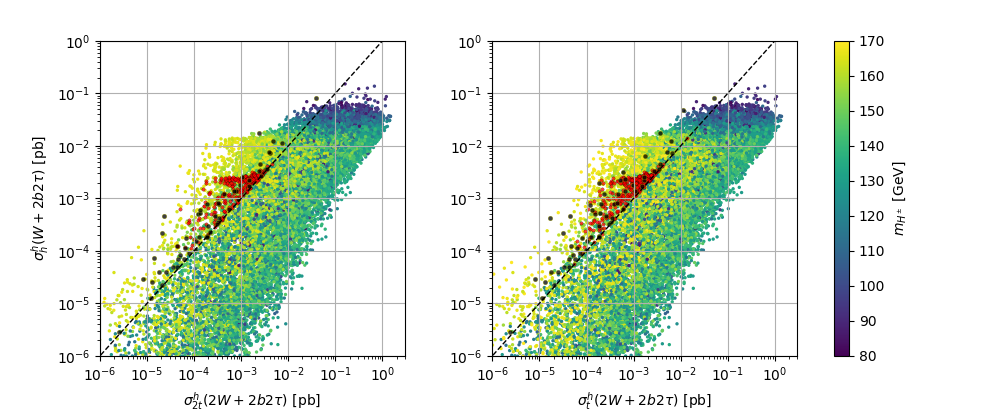}
	\caption{Values of  2 $\sigma(pp \to H^\pm h)\times {\rm BR}(H^\pm \to W^\pm h)\times {\rm BR}(h \to b\bar{b}) \times {\rm BR}(h \to \tau^+\tau^-)$ are compared with those of  2 $\sigma_{t\bar{t}} \times {\rm BR}(t \to b H^+)\times {\rm BR}(\bar{t} \to \bar{b} W^-)\times {\rm BR}(H^\pm \to W^\pm h)\; \times $  ${\rm BR}(h \to \tau^+\tau^-)$ 
		(left) and 
		$ \sigma(pp \to t\bar{b}H^-)\times {\rm BR}(t \to b W+)\times {\rm BR}(H^\pm \to W^\pm h)\; \times $ 
		$ {\rm BR}(h \to \tau^+\tau^-)$ 
		(right). The red points identify the values of  
		$\sigma(H^+ H^-)\times {\rm BR}(H^\pm \to W^\pm h)^2 \times {\rm BR}(h \to b\bar b) \times  
		{\rm BR}(h \to \tau^+\tau^-)$ 
		which also exceeds those of the latter two top mediated processes, respectively. 
		The colour bar denotes the mass of the charged Higgs boson. Results are for the 2HDM Type-I with the exception of the black points, which 
		refer to the 2HDM Type-X rates for $\sigma(pp \to H^\pm h)\times {\rm BR}(H^\pm \to W^\pm h)\times {\rm BR}(h \to \tau^+\tau^-)^2$.} \label{fig4}
\end{figure}
\begin{figure}[hbtp]
	\centering
	\includegraphics[scale=0.65]{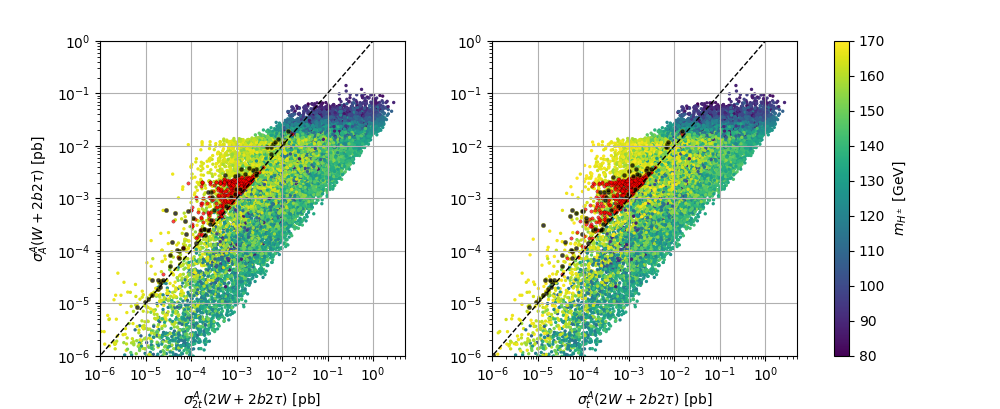}
	\caption{Values of  2 $\sigma(pp \to H^\pm A)\times {\rm BR}(H^\pm \to W^\pm A)\times {\rm BR}(A \to b\bar{b}) \times {\rm BR}(A \to \tau^+\tau^-)$ are compared with those of  2 $\sigma_{t\bar{t}} \times {\rm BR}(t \to b H^+)\times {\rm BR}(\bar{t} \to \bar{b} W^-)\times {\rm BR}(H^\pm \to W^\pm A)\; \times $  ${\rm BR}(A \to \tau^+\tau^-)$ 
		(left) and 
		$ \sigma(pp \to t\bar{b}H^-)\times {\rm BR}(t \to b W^+)\times {\rm BR}(H^\pm \to W^\pm A)\; \times $ 
		$ {\rm BR}(A \to \tau^+\tau^-)$ 
		(right). The red points identify the values of  
		$\sigma(H^+ H^-)\times {\rm BR}(H^\pm \to W^\pm A)^2 \times {\rm BR}(A \to b\bar b) \times  
		{\rm BR}(A \to \tau^+\tau^-)$ 
		which also exceeds those of the latter two top mediated processes, respectively. 
		The colour bar denotes the mass of the charged Higgs boson. Results are for the 2HDM Type-I with the exception of the black points, which 
		refer to the 2HDM Type-X rates for $\sigma(pp \to H^\pm A)\times {\rm BR}(H^\pm \to W^\pm A)\times {\rm BR}(A \to \tau^+\tau^-)^2$.} \label{fig5}
\end{figure}

In Figure \ref{fig2}, again,  for the 2HDM Type-I scenario and the parameter ranges given in Table \ref{param_scan}, we show the relative magnitudes of the production and decay rates of $4b$ final states from $pp \to H^\pm h \to W^{\pm } h h$  and from the two top (anti)quark processes. In the left(right) panel, the cross sections of the di-Higgs process $\sigma^{h}_h(W+4b) $ are compared with those of the process $\sigma_{2t}(2W + 4b)$($\sigma_{t}(2W + 4b)$). The red points shown in these two plots demonstrate cases where also the $pp\to H^+ H^-$ process yielding $\sigma^h_h(2 W + 4b)$ can be larger than the top (anti)quark processes. In Figure  \ref{fig3}, similar to Figure \ref{fig2}, we show the corresponding  results of the production of $4b$ final states from $pp \to H^\pm A\to W^{\pm } A A$. 
It is therefore clear that the combination of charged and neutral current di-Higgs production can provide a significant alternative discovery channel for $H^\pm$ states at the LHC in the 2HDM Type-I.  This conclusion can in fact be extended to the case of the $\sigma^{h}_h(2W+2b2\tau) $ and
the $\sigma^{h}_h(W+2b2\tau) $ signatures versus the  $\sigma_{2t}(2W + 2b2\tau)$  or $\sigma_{t}(2W + 2b2\tau)$ ones, too, as demonstrated in Figures \ref{fig4}--\ref{fig5}. In fact, these last two figures also report 2HDM Type-X rates for $\sigma(pp \to H^\pm h)\times {\rm BR}(H^\pm \to W^\pm h)\times {\rm BR}(h \to \tau^+\tau^-)^2$ (Figure~\ref{fig4})
and $\sigma(pp \to H^\pm A)\times {\rm BR}(H^\pm \to W^\pm A)\times {\rm BR}(A \to \tau^+\tau^-)^2$ (Figure~\ref{fig5}), showing that, for this
Yukawa structure, the $W+4\tau$ signature is the one of interest, as here BR$(h/A \to b\bar b)$ is very small compared to BR$(h/A \to \tau^+\tau^-)$.
Conversely, the $W+4\tau$ signature is of little relevance for the 2HDM Type-I. As for the $2W+4\tau$ case, this is not accessible in either scenario.
Needless to say, 
 for all the parameter space points considered, in both 2HDM Type-I and -X, the decay channels of the charged Higgs boson are dominated by $ \hp \ra W h/A $ with a BR that could reach 100\% in the alignment limit~\cite{Akeroyd:2016ymd,Arhrib:2016wpw,Arhrib:2018ewj}. 

Finally, it should be pointed out that, when a final state is specified, the di-Higgs signature should sum over all processes which can produce such a final state. For example, when $W +4b$ is considered, the cross section from di-Higgs processes should be summed over those of the  intermediate states $W^\pm +h h$, $W^\pm + h A$ (or $W^\pm + A h$)  and $W^\pm + A A$, in turn emerging from the processes $pp \to H^\pm h/A$.  Obviously, the ability to rely on different subchannels all leading to the same final state is a welcome feature when we tension the new discovery modes to the established ones. However, notice that in the `mixed decay chain', i.e., when an intermediate state of the type $W^\pm + h A$ (or $W^\pm + A h$) is produced, from a selection analysis perspective, it would not be possible to enforce an equal mass constraint on the  fermionic systems comprised by two $b\bar b$ pairs or $b\bar b$ and $\tau^+\tau^-$ pairs, so we do expect the experimental efficiency of this case to be less than when  two identical neutral Higgs bosons emerge from the production and decay stages of the di-Higgs processes.  This is why we have refrained from presenting the equivalent of Figures \ref{fig2}--\ref{fig5} for the mixed decay chain.

\section{Benchmark Points}
In order to examine the salient features of the advocated parameter space and to  encourage experimentalists to search for such new discovery channels at the LHC (and/or future colliders), we propose eight Benchmark Points (BPs) for both the 2HDM Type-I and -X scenarios. These BPs are presented in Tables \ref{BPsTypeI} and \ref{BPsTypeX}, respectively. BPs satisfying both conditions $m_{H^\pm} + m_b > m_t$ and $m_{H^\pm} + m_b <  m_t$ are chosen.

Before we list our BPs in detail, we  first show the fractions of parameter space that survive  theoretical and experimental constraints. This information is given in Table \ref{compar_scan} for both Type-I and -X.
\begin{table}[hbtp]
	\centering
	\begin{tabular}{|l|l|l|l|l|} \hline\hline
		& \makecell{Allowed by theoretical \\ constraints} & \makecell{Allowed by \\ \texttt{HiggsBounds}} & \makecell{Allowed by \\ \texttt{HiggsSignals}} & \makecell{Allowed \\ by all constraints}  \\ \hline
		\,Type-I & ~\qquad\quad $5.36$ \% & \quad $17.96$ \% & \quad $55.36$ \% & ~\qquad $0.53$ \% \\ \hline
		Type-X & ~\qquad\quad $5.34$ \% & \quad ~\,$1.37$ \% & \quad ~\,$7.50$ \% & ~\qquad $0.0055$ \% \\ \hline\hline
	\end{tabular}
	\caption{Details of the allowed points in parameter space after imposing theoretical and experimental constraints. Note that EWPO and flavour constraints are essentially always satisfied in both 2HDM Types. Here, we have used a dataset with $5\times10^6$ sample points.} \label{compar_scan}
\end{table} 
In the Type-I scenario, some $0.53\%$ of the points generated in our scan can survive after all constraints have been imposed. In contrast, in the Type-X scenario, only $0.0055\%$ survive.
It is found that the fractions of points that survive the theoretical constraints for these two scenarios are almost equal, which is understandable since they  share the same Higgs potential.  Compared with the Type-I scenario, in Type-X, fewer points pass  \texttt{HiggsBounds} constraints enforcing limits from nil searches for additional Higgs bosons (beyond the SM-like one) performed at LEP, Tevatron and the LHC. Similarly, for the Type-X scenario, again, fewer points than in Type-I  pass the constraints from the SM-like Higgs boson data gathered by the LHC experiments and included in  \texttt{HiggsSignals}. The reason can be attributed to the fact that, when the couplings to quarks are suppressed and those  to leptons are enhanced, it is crucial to predict the correct production rate for $gg \to H$ and the correct decay fractions for $H \to b \bar{b}$ and $H \to \gamma \gamma$ and, in the light of the couplings in Table \ref{Yukawa-2}, this is more easily achieved in Type-I than -X. 

For the BPs given in Tables \ref{BPsTypeI} and \ref{BPsTypeX}, it is observed that the mass splittings between the two lightest neutral Higgs bosons ($h$ and $A$) are not allowed to be too large, which is also due to the constraints from EWPOs. It should also be emphasised that these BPs have relatively light neutral Higgs bosons, which can also be searched for at  future lepton colliders, like the 240 GeV `Higgs factories' of CEPC, FCC-ee and ILC, where the search modes would involve  four fermion final states, like $4b$, $2b2\tau$ and $4\tau$, while at the LHC such signal events might be overwhelmingly concealed by the huge QCD background\footnote{In this sense, the triggering possibilities (e.g., by exploiting $W^\pm\to l^\pm \nu_l$ decays, with $l=e,\mu$) in our advocated LHC signatures offered by the presence of one or two $W^\pm$'s are crucial to establishing the corresponding signatures,  involving such four fermion combinations.}. Finally, notice that, for all our BPs, to compute the production and decay rates, we have taken $m_H=125$ GeV, $m_t=172.5$ GeV, $m_b=4.75$ GeV and $m_\tau=1.77684$ GeV.

\subsection{Benchmark Points for Type-I}

In Table \ref{BPsTypeI}, we present eight BPs for the 2HDM Type-I scenario.  Here, BP1,2,3,4 and 5 satisfy the condition $m_{H^\pm} + m_b < m_t$ while the rest of the points satisfy the condition $m_{H^\pm}+ m_b > m_t$.  For every BP in the table, we provide the cross sections of the $W + 4b$ and $W + 2b2\tau$ signatures from both di-Higgs and top (anti)quark processes, to gauge the relative yields against alternative numerical estimates. Let us now discuss the features of each BP.
There are two observations to make regarding  BP1. 
\begin{enumerate}
\item The total production rate of $2W + 4b$ from top (anti)quark processes is around $5$ fb\footnote{Here, we refer to the $q\bar q,gg\to t\bar b H^-$ + c.c. rates, as, to remind the reader, the $q\bar q,gg\to t\bar t\to t \bar b H^-$ case  is provided only for reference purposes, as this channel  is what is typically being used in experimental analyses for the case $m_{H^\pm}<m_t$.}, while the total production rate of $2W + 4b$ from Higgs pair production of charged Higgs bosons is $\approx 23.1$ fb. Furthermore, the production rate of $W+4b$ from di-Higgs (charged current) production channels can reach to $174$ fb, which is 7 times larger than the final state of $2W + 4b$ from the di-Higgs (neutral current) production channels. 

\item While for the total production rate of $2W+2b2\tau$ the top (anti)quark processes can yield only $0.46$ fb, pair production of charged Higgs boson bosons yields $2.13$ fb. Furthermore, the production rate of $W + 2b 2\tau$ from di-Higgs processes can even reach $16$ fb.
\end{enumerate}
Such two observations also hold for BP2,3 and 5. For BP4,6 and 8, the charged Higgs bosons are so heavy that a top (anti)quark cannot decay into a charged Higgs boson plus a $b$ (anti)quark. Therefore, the corresponding $\sigma_{2t}(2W + 4 b)$ rates vanish due to this kinematic reason while $\sigma_{t}(2W + 4 b)$ is
generally rather small, below the fb level. For BP4, the production rate of $2W + 4b$ from  top (anti)quark processes yields a cross section of $3.88$ fb while that from Higgs-pair production involving charged Higgs bosons yields $20.24$ fb and that of $W + 4b$ yields $169.35$ fb.
For the final state $2W+2b2\tau$, the production rate from top (anti)quark processes is $0.36$ fb while that from pair production of charged Higgs bosons is $1.87$ fb. Finally, the signature $W + 2b 2\tau$ from di-Higgs channels is $15.63$ fb. (Such features are similar to those of BP6, 7 and 8.)

\begin{table}[hbtp]
	\centering
	\begin{tabular}{|l|l|l|l|l|l|l|l|l|}\hline\hline
		~~\, Parameters & ~BP1 & ~BP2   & ~BP3 & ~BP4 & ~BP5 & ~BP6 & ~BP7  & ~BP8\\\hline
		~\quad $m_h$ (GeV) &  ~$91.00$ & ~$96.84$ & $103.34$ & ~$99.61$ & ~$95.57$ & ~$94.00$ & ~$94.00$ & ~$94.00$\\
		~\quad $m_H$ (GeV) &  $125.00$ & $125.00$  & $125.00$  & $125.00$ & $125.00$  & $125.00$ & $125.00$ & $125.00$\\
		~\quad $m_A$ (GeV) &  $102.04$ & $112.35$  & ~$93.80$  & ~$88.98$ & ~$94.41$  & $105.00$ & $105.00$ & $105.00$\\  
		~\quad $m_{H^\pm}$ (GeV) &  $167.02$ & $166.34$  & $161.02$  & $169.46$ & $167.02$ & $176.00$ & $186.00$ & $196.00$\\ 
		~~\qquad $s_{\beta-\alpha}$ & $-0.18$ & $-0.11$ & $-0.19$  & $-0.06$ & $-0.09$ & $-0.09$ & $-0.09$ & $-0.09$\\ 
		~~\qquad $\tan\beta$ & ~$40.87$ & ~$58.17$ & ~$54.79$ & ~$39.10$ & ~$32.44$ & ~$30.00$ & ~$30.00$ & ~$30.00$\\
		~\quad $m_{12}^2$ (GeV$^2$) & $204.22$ & $161.85$ & $196.73$  & $252.94$ & $277.81$ & $294.00$ & $294.00$ & $294.00$ \\ \hline\hline
		
		~~ $\sigma^{h}_{2t}(2W + 4b)$ & ~$2.30$ & ~$1.65$ & ~$2.06$ & \,~~$-$ & ~$2.42$ & \,~~$-$ & \,~~$-$ & \,~~$-$\\
		~~ $\sigma^{h}_{t}(2W+4b)$ & ~$3.85$ & ~$2.35$ & ~$2.26$ & ~$0.85$ & ~$3.84$ & ~$5.03$ & ~$4.68$ & ~$3.52$\\ 
		~~ $\sigma^{A}_{2t}(2 W + 4b)$ & ~$0.70$ & ~$0.25$ & ~$4.63$ & \,~~$-$ &  ~$2.47$ & \,~~$-$ & \,~~$-$ & \,~~$-$\\
		~~ $\sigma^{A}_{t}(2W+4b)$ & ~$1.17$ & ~$0.36$ & ~$5.07$ & ~$3.03$ &  ~$3.92$ & ~$0.83$ & ~$0.44$ & ~$1.08$\\ \hline \hline
		
		\quad$\sigma^{h}_{h}(2W+4b)$ & $13.58$ & $15.99$ & ~$2.29$ & ~$0.97$ & ~$5.38$ & $14.08$ & $13.27$ & ~$7.35$\\	
		\quad$\sigma^{h}_{A}(2W+4b)$ & ~$4.13$ & ~$2.44$ & ~$5.14$ & ~$3.46$ & ~$5.50$ & ~$2.32$ & ~$1.25$ & ~$2.24$\\
		\quad$\sigma^{A}_{A}(2W+4b)$ & ~$1.26$ & ~$0.37$ & $11.55$ & $12.35$ & ~$5.62$ & ~$0.38$ & ~$0.12$ & ~$0.68$\\
		\quad$\sigma^{A}_{h}(2W+4b)$ & ~$4.13$ & ~$2.44$ & ~$5.14$ & ~$3.46$ & ~$5.50$ & ~$2.32$ & ~$1.25$ & ~$2.24$\\ \hline\hline
		\,\quad$\sigma^{h}_{h}(W+4b)$ & $75.88$ & $77.61$ & $26.47$  & $17.68$ & $46.00$ & $73.25$ & $68.00$ & $48.81$\\
		\,\quad$\sigma^{h}_{A}(W+4b)$ & $23.07$ & $11.86$ & $59.44$  & $63.04$ & $47.00$ & $12.07$ & ~$6.42$ & $14.90$\\		
		\,\quad$\sigma^{A}_{A}(W+4b)$ & $17.48$ & ~$6.12$ & $64.39$ & $69.22$ & $43.51$ & ~$9.16$ & ~$4.91$ & $11.45$\\ 
		\,\quad$\sigma^{A}_{h}(W+4b)$ & $57.51$ & $40.06$ & $28.68$ & $19.41$ & $42.59$ & $55.59$ & $52.02$ & $37.51$\\ \hline\hline	
		~\,$\sigma^{h}_{2t}(2W+2b2\tau)$ & ~$0.21$ & ~$0.15$ & ~$0.19$ & ~~$-$ & ~$0.22$ & ~~$-$ & ~~$-$ & ~~$-$\\
		~\,$\sigma^{h}_{t}(2W+2b2 \tau)$ & ~$0.35$ & ~$0.22$ & ~$0.21$ & ~$0.08$ & ~$0.35$ & ~$0.46$ & ~$0.43$ & ~$0.32$\\ 
		~\,$\sigma^{A}_{2t}(2W+2b2\tau)$ & ~$0.07$ & ~$0.02$ & ~$0.43$ & ~~$-$ &  ~$0.23$ & ~~$-$ & ~~$-$ & ~~$-$\\
		~\,$\sigma^{A}_{t}(2W + 2b2\tau)$ & ~$0.11$ & ~$0.03$ & ~$0.47$ & ~$0.28$ &  ~$0.37$ & ~$0.08$ & ~$0.04$ & ~$0.10$\\ \hline	\hline
		~\,$\sigma^{h}_{h}(2W+2b2\tau)$ & ~$1.24$ & ~$1.48$ & ~$0.21$ & ~$0.09$ & ~$0.50$  & ~$1.29$ & ~$1.22$ & ~$0.67$\\	
		~\,$\sigma^{h}_{A}(2W+2b2\tau)$ & ~$0.38$ & ~$0.23$ & ~$0.48$ & ~$0.32$ & ~$0.51$  & ~$0.21$ & ~$0.11$ & ~$0.21$\\ 
		~\,$\sigma^{A}_{A}(2W+2b2\tau)$ & ~$0.12$ & ~$0.04$ & ~$1.08$ & ~$1.14$ & ~$0.52$ & ~$0.04$ & ~$0.01$ & ~$0.07$\\
		~\,$\sigma^{A}_{h}(2W+2b2\tau)$ & ~$0.39$ & ~$0.24$ & ~$0.48$ & ~$0.32$ & ~$0.51$ & ~$0.22$ & ~$0.12$ & ~$0.21$\\ \hline  \hline
		~~$\sigma^{h}_{h}(W+2b2\tau)$ & ~$6.93$ & ~$7.17$ & ~$2.47$  & ~$1.64$ & ~$4.24$ & ~$6.73$ & ~$6.24$ & ~$4.48$\\
		~~$\sigma^{h}_{A}(W+2b2\tau)$ & ~$2.18$ & ~$1.14$ & ~$5.54$  & ~$5.81$ & ~$4.38$ & ~$1.15$ & ~$0.61$ & ~$1.42$\\ 	
		~~$\sigma^{A}_{A}(W+2b2\tau)$ & ~$1.66$ & ~$0.59$ & ~$6.00$ & ~$6.38$ & ~$4.06$ & ~$0.87$ & ~$0.47$ & ~$1.09$\\ 
		~~$\sigma^{A}_{h}(W+2b2\tau)$ & ~$5.25$ & ~$3.70$ & ~$2.68$ & ~$1.80$ & ~$3.92$ & ~$5.10$ & ~$4.78$ & ~$3.44$\\ \hline \hline 
	\end{tabular}
	\caption{Mass spectra and mixing angles in the Type-I scenario, alongside cross sections (in fb) from different production channels, are shown. } \label{BPsTypeI}
\end{table}


\subsection{Benchmark Points for Type-X}

In Table \ref{BPsTypeX}, eight BPs for the Type-X scenario are presented. In this scenario, the light neutral Higgs bosons dominantly decay into  $\tau$'s due to the structure and size of the  Yukawa couplings, as seen in Table \ref{Yukawa-2}.  Here, for the advocated new discovery channels,  we only tabulate the results for $W + 4 \tau$ final states, since (as previously shown) this is the only signature available. Specifically, notice that the only viable channel, in this Type, is the associated production of a charged Higgs boson with a neutral Higgs, $h$ or $A$, i.e., $pp \to H^\pm h/A$, which would lead to a $W^\pm + 4\tau$ final state.  
For BP1,3 and 4, the condition $m_{H^\pm} + m_b < m_t$ is satisfied. For BP2,5,6,7 and 8, the condition $m_{H^\pm} + m_b > m_t$ is satisfied (so that charged Higgs boson production from top (anti)quark decay is forbidden).
For BP1, the total production rate of $2W + 2b2\tau$ from top (anti)quark channels is $3.72$ fb and that of $W + 4 \tau$ from di-Higgs production is $30.33$ fb, around 6 times larger. Similarly for BP3 and BP4. For BP2, the total production rate of $2W+2b2\tau$ from top (anti)quark  channels is $11.62$ fb and that of $W + 4\tau$ from the di-Higgs production channels is $85.82$ fb. Hence, the $W+4\tau$ signature from the aforementioned di-Higgs model is always more sizeable than the $2W+2b2\tau$ one from top (anti)quark process dynamics. 

\begin{table}[hbtp]
	\centering
	\begin{tabular}{|l|l|l|l|l|l|l|l|l|}\hline\hline
		~~\, Parameters & ~BP1 & ~BP2   & ~BP3 & ~BP4 & ~BP5 & ~BP6 & ~BP7  & ~BP8\\\hline
		~\quad $m_h$ (GeV) &  ~$83.66$ & ~$83.23$ & $100.04$ & $115.35$ & ~$95.12$ & ~$84.84$ & $103.41$ & ~$86.87$\\
		~\quad $m_H$ (GeV)&  $125.00$ & $125.00$  & $125.00$  & $125.00$ & $125.00$  & $125.00$ & $125.00$ & $125.00$\\
		~\quad $m_A$ (GeV)&  $113.60$ & $109.52$  & ~$93.55$  & ~$79.30$ & $101.38$  & $108.83$ & ~$90.46$ & $112.97$\\  
		~\quad $m_{H^\pm}$ (GeV)&  $166.22$ & $169.14$  & $166.18$  & $158.67$ & $169.99$ & $176.64$ & $186.78$ & $195.68$\\ 
		~~\qquad $s_{\beta-\alpha}$ & $-0.10$ & $-0.13$ & $-0.17$  & $-0.10$ & $-0.13$ & $-0.12$ & $-0.13$ & $-0.12$\\ 
		~~\qquad $\tan\beta$ & ~$18.57$ & ~$14.41$ & ~$10.51$ & ~$17.42$ & ~$13.90$ & ~$15.37$ & ~$15.36$ & ~$14.53$\\
		~\quad $m_{12}^2$ (GeV$^2$)& $367.17$ & $408.42$ & $801.13$  & $728.57$ & $645.95$ & $437.53$ & $631.00$ & $456.00$ \\ \hline\hline
		~$\sigma^{h}_{2t} (2W+2b2\tau)$ & ~$2.42$ & \,~~$-$ & ~$2.54$ & ~$0.18$ & \,~~$-$ & \,~~$-$ & \,~~$-$ & \,~~$-$\\
		~$\sigma^{h}_{t} (2W+2b2\tau)$ & ~$3.61$ & $11.33$ & ~$3.68$ & ~$0.19$ & ~$1.77$ & $13.23$ & ~$1.84$ & $19.07$\\
		~$\sigma^{A}_{2t}(2W + 2b 2\tau) $ & ~$0.08$ & \,~~$-$ & ~$4.74$ & ~$6.43$ & \,~~$-$ & \,~~$-$ & \,~~$-$ & \,~~$-$\\
		~$\sigma^{A}_{t}(2W+2b2\tau)$ & ~$0.11$ & ~$0.29$ & ~$6.88$ & ~$6.72$ & ~$0.99$ & ~$0.30$ & $13.35$ & ~$0.97$\\ \hline \hline
		\quad$\sigma^{h}_{h}(W+4\tau)$ & $17.29$ & $48.02$ & ~$4.48$  & ~$0.19$ & ~$6.29$ & $66.34$ & ~$7.55$ & $80.44$\\
		\quad$\sigma^{h}_{A}(W+4\tau)$ & ~$0.54$ & ~$1.25$ & ~$8.36$  & ~$6.72$ & ~$3.51$ & ~$1.48$ & $54.84$ & ~$4.07$\\
		\quad$\sigma^{A}_{A}(W+4\tau)$ & ~$0.38$ & ~$0.93$ & ~$9.33$  & $10.68$ & ~$3.31$ & ~$1.14$ & $64.36$ & ~$3.14$\\
		\quad$\sigma^{A}_{h}(W+4\tau)$ & $12.12$ & $35.62$ & ~$5.00$  & ~$0.29$ & ~$5.94$ & $50.98$ & ~$8.86$ & $62.04$\\ \hline\hline		  
	\end{tabular}
	\caption{Mass spectra and mixing angles in the Type-X scenario, alongside cross sections (in fb) from different production channels, are shown.} \label{BPsTypeX}
\end{table}

\section{Conclusions}
By adding a Higgs doublet to the SM, one can address some fundamental issues of the latter, like the structure  of EWSB, the flavour puzzle, the need for   additional CP violation, the dynamics of EW baryogenesis, the provision of suitable Dark Matter (DM)  candidates and so on. This addition generates a 2HDM, which has
been the subject of extensive studies in the literature. Here, we wanted to examine the phenomenology of the 2HDM in regions of its parameter space that can offer us new ways of detecting the charged Higgs boson state emerging in this BSM construct, above and beyond what is already established in such a literature.

Specifically, the experimental teams at the LHC are presently searching for $H^\pm$ states in 2HDM realisations by assuming these to be produced in association with top (anti)quarks. The advantage of exploiting such production processes is twofold. On the one hand, (anti)top quarks are produced (in pairs) via QCD interactions, which thus have no BSM parameter dependence. On the other hand, being QCD induced, the corresponding cross sections at the LHC are rather large. However, when it comes to the top (anti)quark interacting with the charged Higgs bosons, via a $H^+\bar t  b$ vertex, a BSM parameter dependence becomes manifest. This can be ascribed to the different ways the two Higgs doublets of this BSM scenario can give mass to elementary fermions, quarks and leptons, via Yukawa couplings. There are typically four possible choices for doing so, referred to as Types. Here, we have been concerned with  two of these,  known as Type-I and -X. In these two realisations of the 2HDM, the $H^\pm$ production and decay rates associated with processes involving   
top (anti)quarks can be surpassed by those in which charged Higgs bosons are produced via EW interactions which only gauge dependent (i.e., they do not depend on 2HDM parameters), when the $H^\pm$ states are searched for through their $H^\pm\to W^{\pm }h/A$ decays and  the SM-like Higgs boson discovered at the LHC is identified with the heavy CP-even state of the 2HDM, $H$, so that the light CP-even one, $h$,  and CP-odd one, $A$, are lighter in comparison. 

Specifically, in this work, we have investigated charged Higgs boson production via $pp\to H^\pm \phi$  ($\phi = h, A$) 
and $pp\to H^+ H^-$ at the LHC with $\sqrt{s} = 14$ TeV in the 2HDM Type-I and -X, after taking into account all up-to-date theoretical and experimental 
constraints. Using these production mechanisms and allowing for  $H^\pm\to W^{\pm }h/A$ decays in all possible combinations, 
we have examined the final states $2W+4b$, $2W+2\tau 2b$, $W+4b$, $W+2\tau 2b$ and $W + 4 \tau$ as potential discovery channels. We have thus compared their yield against that of the  channels involving top (anti)quarks and shown that the former can be larger than the latter for a large interval of $m_{H^\pm}$ values, up to 197 GeV or so when $\tan\beta$ is large. However, the most interesting region of 2HDM parameter space in both Types studied here is the one where $m_{H^\pm}\lsim m_t$, as it gives potentially detectable event rates for most of the aforementioned final states. Thus, in order to enable the ATLAS and CMS Collaborations to confirm or disprove this possibility, we have finally presented a variety of BPs in both  2HDM Type-I and -X amenable to experimental investigation.

\section*{Acknowledgements}
The work of AA, RB, MK and BM is supported by the Moroccan Ministry of Higher Education and Scientific Research MESRSFC and
CNRST Project PPR/2015/6. The work of SM is supported in part through the NExT Institute and the STFC Consolidated Grant No. ST/L000296/1.
Y. W. is supported by the `Scientific Research Funding Project for
Introduced High-level Talents' of the Inner Mongolia Normal
University Grant No. 2019YJRC001.  Q.-S. Yan's work is supported by the Natural Science Foundation of China Grant No. 11875260.

\appendix


\begin{thebibliography}{}


\bibitem{Aad:2012tfa}
G.~Aad \textit{et al.} [ATLAS],
Phys. Lett. B \textbf{716} (2012)  1,
[arXiv:1207.7214 [hep-ex]].

\bibitem{Chatrchyan:2012ufa}
S.~Chatrchyan \textit{et al.} [CMS],
Phys. Lett. B \textbf{716} (2012)  30,
[arXiv:1207.7235 [hep-ex]].

\bibitem{Aad:2019mbh}
G.~Aad \textit{et al.} [ATLAS],
Phys. Rev. D \textbf{101} (2020), no. 1 012002,
[arXiv:1909.02845 [hep-ex]].

\bibitem{Sirunyan:2018koj}
A.~M.~Sirunyan \textit{et al.} [CMS],
Eur. Phys. J. C \textbf{79} (2019), no. 5 421,
[arXiv:1809.10733 [hep-ex]].



\bibitem{Barger:1993th}
V.~D.~Barger, R.~J.~N.~Phillips and D.~P.~Roy,
Phys. Lett. B \textbf{324} (1994)  236,
[arXiv:hep-ph/9311372 [hep-ph]]

\bibitem{Gunion:1986pe}
J.~F.~Gunion, H.~E.~Haber, F.~E.~Paige, W.~K.~Tung and S.~S.~D.~Willenbrock,
Nucl. Phys. B \textbf{294} (1987)  621.

\bibitem{Barnett:1987jw}
R.~M.~Barnett, H.~E.~Haber and D.~E.~Soper,
Nucl. Phys. B \textbf{306} (1988)  697.

\bibitem{DiazCruz:1992gg}
J.~L.~Diaz-Cruz and O.~A.~Sampayo,
Phys. Rev. D \textbf{50} (1994)  6820.

\bibitem{Moretti:1999bw}
  S.~Moretti and D.~P.~Roy,
  Phys.\ Lett.\ B {\bf 470} (1999) 209,
  [hep-ph/9909435].

\bibitem{Miller:1999bm}
  D.~J.~Miller, S.~Moretti, D.~P.~Roy and W.~J.~Stirling,
  Phys.\ Rev.\ D {\bf 61} (2000) 055011,
  [hep-ph/9906230].

\bibitem{Guchait:2001pi}
M.~Guchait and S.~Moretti,
JHEP \textbf{01} (2002)  001,
[hep-ph/0110020].

\bibitem{Alwall:2003tc}
  J.~Alwall, C.~Biscarat, S.~Moretti, J.~Rathsman and A.~Sopczak,
  Eur.\ Phys.\ J.\ C {\bf 39S1} (2005) 37,
  [hep-ph/0312301].

\bibitem{Alwall:2004xw}
  J.~Alwall and J.~Rathsman,
  JHEP {\bf 0412} (2004) 050,
  [hep-ph/0409094].

\bibitem{BarrientosBendezu:1998gd}
  A.~A.~Barrientos Bendezu and B.~A.~Kniehl,
  Phys.\ Rev.\ D {\bf 59} (1999) 015009,
  [hep-ph/9807480].

  
\bibitem{Moretti:1998xq}
  S.~Moretti and K.~Odagiri,
  Phys.\ Rev.\ D {\bf 59} (1999) 055008,
  [hep-ph/9809244].

\bibitem{BarrientosBendezu:1999vd}
  A.~A.~Barrientos Bendezu and B.~A.~Kniehl,
  Phys.\ Rev.\ D {\bf 61} (2000) 097701,
  [hep-ph/9909502].

\bibitem{Moretti:1996ra}
S.~Moretti and K.~Odagiri,
Phys.\ Rev.\ D {\bf 55} (1997) 5627,
[hep-ph/9611374].

\bibitem{Arhrib:2015gra}
A.~Arhrib, K.~Cheung, J.~S.~Lee and C.~T.~Lu,
JHEP \textbf{05} (2016)  093,
[arXiv:1509.00978 [hep-ph]].

\bibitem{HernandezSanchez:2012eg}
J.~Hernandez-Sanchez, S.~Moretti, R.~Noriega-Papaqui and A.~Rosado,
JHEP {\bf 1307} (2013) 044,
[arXiv:1212.6818 [hep-ph]].

\bibitem{Hernandez-Sanchez:2020vax}
J.~Hernández-Sánchez, C.~G.~Honorato, S.~Moretti and S.~Rosado-Navarro,
Phys.\ Rev.\ D {\bf 102} (2020), no. 5  055008,
[arXiv:2003.06263 [hep-ph]].

\bibitem{Dittmaier:2007uw}
S.~Dittmaier, G.~Hiller, T.~Plehn and M.~Spannowsky,
Phys. Rev. D \textbf{77} (2008)  115001,
[arXiv:0708.0940 [hep-ph]].

\bibitem{cpyuan} S. Kanemura and C. Yuan,
Phys. Lett. B {\bf 530} (2002)  188, 
[arXiv:hep-ph/0112165].

\bibitem{Enberg:2014pua}
R.~Enberg, W.~Klemm, S.~Moretti, S.~Munir and G.~Wouda,
Nucl. Phys. B \textbf{893} (2015)  420,
[arXiv:1412.5814 [hep-ph]].

\bibitem{Enberg:2015qsa}
R.~Enberg, W.~Klemm, S.~Moretti, S.~Munir and G.~Wouda,
[arXiv:1506.04409 [hep-ph]].

\bibitem{Enberg:2017gyo}
R.~Enberg, W.~Klemm, S.~Moretti and S.~Munir,
PoS ICHEP {\bf 2016} (2017) 1174,
[arXiv:1704.06405 [hep-ph]].

\bibitem{Enberg:2018nfv}
R.~Enberg, W.~Klemm, S.~Moretti and S.~Munir,
PoS CORFU {\bf 2018} (2018) 013,
[arXiv:1812.08623 [hep-ph]].

\bibitem{Enberg:2018pye}
R.~Enberg, W.~Klemm, S.~Moretti and S.~Munir,
Eur.\ Phys.\ J.\ C {\bf 79} (2019), no. 6  512,
[arXiv:1812.01147 [hep-ph]].

\bibitem{BarrientosBendezu:1999gp}
A.~A.~Barrientos Bendezu and B.~A.~Kniehl,
Nucl. Phys. B \textbf{568} (2000) 305,
[arXiv:hep-ph/9908385 [hep-ph]].

\bibitem{Brein:1999sy}
O.~Brein and W.~Hollik,
Eur. Phys. J. C \textbf{13} (2000) 175,
[arXiv:hep-ph/9908529 [hep-ph]].

\bibitem{Moretti:2001pp}
S.~Moretti,
J.\ Phys.\ G {\bf 28} (2002) 2567,
[hep-ph/0102116].

\bibitem{Moretti:2003px}
S.~Moretti and J.~Rathsman,
Eur.\ Phys.\ J.\ C {\bf 33} (2004) 41,
[hep-ph/0308215].

\bibitem{Aoki:2011wd}
M.~Aoki, R.~Guedes, S.~Kanemura, S.~Moretti, R.~Santos and K.~Yagyu,
Phys. Rev. D \textbf{84} (2011)  055028,
[arXiv:1104.3178 [hep-ph]].


\bibitem{Akeroyd:2016ymd}
A.~G.~Akeroyd, M.~Aoki, A.~Arhrib, L.~Basso, I.~F.~Ginzburg, R.~Guedes, J.~Hernandez-Sanchez, K.~Huitu, T.~Hurth and M.~Kadastik, \textit{et al.}
Eur. Phys. J. C \textbf{77} (2017), no. 5 276,
[arXiv:1607.01320 [hep-ph]].
		
\bibitem{Arhrib:2018ewj}
A.~Arhrib, R.~Benbrik, H.~Harouiz, S.~Moretti and A.~Rouchad,
[arXiv:1810.09106 [hep-ph]].

\bibitem{Coleppa:2019cul}
B.~Coleppa, A.~Sarkar and S.~K.~Rai,
Phys. Rev. D \textbf{101} (2020), no. 5 055030,
[arXiv:1909.11992 [hep-ph]].

\bibitem{Akeroyd:1998dt}
A. G. Akeroyd,
Nucl. Phys. B 544 (1999)  557,
[arXiv:hep-ph/9806337 [hep-ph]].

\bibitem{Arhrib:2016wpw}
A.~Arhrib, R.~Benbrik and S.~Moretti,
Eur. Phys. J. C \textbf{77} (2017), no. 9 621,
[arXiv:1607.02402 [hep-ph]].

\bibitem{Abazov:2009ae}
V.~M.~Abazov \textit{et al.} [D0],
Phys. Rev. D \textbf{80} (2009)  071102,
[arXiv:0903.5525 [hep-ex]].

\bibitem{Abulencia:2005jd}
A.~Abulencia \textit{et al.} [CDF],
Phys. Rev. Lett. \textbf{96} (2006)  042003,
[arXiv:hep-ex/0510065 [hep-ex]].

\bibitem{Abbiendi:2013hk}
G.~Abbiendi \textit{et al.} [ALEPH, DELPHI, L3, OPAL and LEP],
Eur. Phys. J. C \textbf{73} (2013)  2463,
[arXiv:1301.6065 [hep-ex]].

\bibitem{Aad:2014kga}
G.~Aad {\it et al.} [ATLAS Collaboration],
JHEP {\bf 1503} (2015)  088,
[arXiv:1412.6663 [hep-ex]].

	\bibitem{Khachatryan:2015qxa}
V.~Khachatryan {\it et al.} [CMS Collaboration],
JHEP {\bf 1511} (2015)  018,
[arXiv:1508.07774 [hep-ex]].

\bibitem{CMS:2016szv}
CMS Collaboration,
CMS-PAS-HIG-16-031.

\bibitem{Aaboud:2018gjj}
M.~Aaboud \textit{et al.} [ATLAS],
JHEP \textbf{09} (2018)  139,
[arXiv:1807.07915 [hep-ex]].

\bibitem{ATLAS:2011yia}
ATLAS Collaboration,
ATLAS-CONF-2011-094.

\bibitem{Sirunyan:2020aln}
A.~M.~Sirunyan \textit{et al.} [CMS],
Phys. Rev. D \textbf{102} (2020), no. 7 072001,
[arXiv:2005.08900 [hep-ex]].

	\bibitem{Tanabashi}
Tanabashi {\it et al.}, 
Phys. Rev. D. {\bf 98} (3)  030001.  

\bibitem{Sirunyan:2019hkq}
A.~M.~Sirunyan \textit{et al.} [CMS],
JHEP \textbf{07} (2019)  142,
[arXiv:1903.04560 [hep-ex]].

\bibitem{Branco:2011iw}
G.~C.~Branco, P.~M.~Ferreira, L.~Lavoura, M.~N.~Rebelo, M.~Sher and J.~P.~Silva,
Phys. Rept. \textbf{516} (2012)  1,
[arXiv:1106.0034 [hep-ph]].

\bibitem{Glashow:1976nt}
S.~L.~Glashow and S.~Weinberg,
Phys.\ Rev.\ D {\bf 15} (1977) 1958.


\bibitem{sta} N. G. Deshpande and E. Ma,
 Phys. Rev. D {\bf 18} (1978) 2574.

\bibitem{uni1}  S. Kanemura,  T. Kubota  and  E. Takasugi, Phys. Lett. B {\bf 313} (1993) 155, [arXiv:hep-ph/9303263 [hep-ph]]. 

\bibitem{uni2} A. G. Akeroyd, A. Arhrib and E. M. Naimi, Phys. Lett. B {\bf 490} (2000) 119, [arXiv:hep-ph/0006035 [hep-ph]]. A. Arhrib, [arXiv:hep-ph/0012353 [hep-ph]].

\bibitem{peskin} M. E. Peskin and T. Takeuchi, Phys. Rev. D {\bf46} (1992) 381.

\bibitem{stu-2HDM1} H.-J. He, N. Polonsky and S. -f. Su, Phys. Rev. D {\bf 64} (2001) 053004, [hep-
ph/0102144].

\bibitem{stu-2HDM2}
 W. Grimus, L. Lavoura, O. M. Ogreid and P. Osland, Nucl. Phys.
B {\bf 801} (2008) 81, [arXiv:0802.4353 [hep-ph]].

\bibitem{stu-2HDM3}
H. E. Haber and D. O’Neil, Phys. Rev.
D {\bf 83} (2011) 055017, [arXiv:1011.6188 [hep-ph]].

\bibitem{HB} P. Bechtle, D. Dercks, S. Heinemeyer, T. Klingl, T. Stefaniak, G. Weiglein, and J. Wittbrodt,
 Eur. Phys. J. C {\bf 80} (2020), no. 12 1211, [arXiv:2006.06007 [hep-ph]].

\bibitem{HS} P. Bechtle, S. Heinemeyer, T. Klingl, T. Stefaniak, G. Weiglein, and J. Wittbrodt,
 Eur. Phys. J. C {\bf 81} (2021), no. 2 145, [arXiv:2012.09197[hep-ph]].

\bibitem{superIso} F. Mahmoudi, Comput. Phys. Commun. {\bf 180} (2009) 1579,
 [arXiv:0808.3144 [hep-ph]].

\bibitem{Bphys1} HFLAV, Y. Amhis \textit{et al.}, Eur. Phys. J. C {\bf 77} (2017) 895,
 [arXiv:1612.07233 [hep-ex]].

\bibitem{Bphys2} LHCb, R. Aaij \textit{et al.}, Phys. Rev. Lett. {\bf 118} (2017) 191801,
 [arXiv:1703.05747 [hep-ex]].

\bibitem{2HDMC} D. Eriksson, J. Rathsman and O. Stal,
 Comput. Phys. Commun. {\bf 181} (2010) 189, [arXiv:0902.0851 [hep-ph]].

\bibitem{Zhu:2001nt}
S.~h.~Zhu,
Phys.\ Rev.\ D {\bf 67} (2003) 075006,
[hep-ph/0112109].

\bibitem{Plehn:2002vy}
T.~Plehn,
Phys.\ Rev.\ D {\bf 67} (2003) 014018,
[hep-ph/0206121].

\bibitem{Berger:2003sm}
E.~L.~Berger, T.~Han, J.~Jiang and T.~Plehn,
Phys.\ Rev.\ D {\bf 71} (2005) 115012,
[hep-ph/0312286].
	
\bibitem{Assamagan:2004gv}
K.~A.~Assamagan, M.~Guchait and S.~Moretti,
[hep-ph/0402057].
	
\bibitem{Stelzer:1994ta}
  T.~Stelzer and W.~F.~Long,
  Comput.\ Phys.\ Commun.\  {\bf 81} (1994) 357,
  [hep-ph/9401258].

\bibitem{PlothowBesch:1992qj}
  H.~Plothow-Besch,
  Comput.\ Phys.\ Commun.\  {\bf 75} (1993) 396.

\end{thebibliography}
\end{document}